\documentclass[prl,twocolumn,floatfix]{revtex4}

\newcommand{\sub}[1]{\mbox{$_{\mbox{\tiny #1}}$}}

\newcommand{\be}{\begin{eqnarray}}
\newcommand{\ee}{\end{eqnarray}}
\newcommand{\ket}[1]{\ensuremath{\left| {#1} \right>}}

\newcommand{\nbar}{\bar{n}}

\newcommand{\caf}{\ensuremath{^{40}{\rm Ca}^{+}\, }}
\newcommand{\cafthree}{\ensuremath{^{43}{\rm Ca}^{+} \,}}

\usepackage{graphicx}
\usepackage{amsmath}
\usepackage{amssymb}

\begin{document}

\title{Memory coherence of a sympathetically cooled trapped-ion qubit}

\author{J. P. Home}
\author{M. J. McDonnell}
\author{D. J. Szwer}
\author{B. C. Keitch}
\author{D. M. Lucas}
\author{D. N. Stacey}
\author{A. M. Steane}

\affiliation{Department of Physics, University of Oxford, Clarendon Laboratory, Parks Road, Oxford OX1 3PU, U.K.}

\date{6 October 2008}

\begin{abstract}
We demonstrate sympathetic cooling of a \cafthree\ trapped-ion ``memory'' qubit by a \caf\ ``coolant'' ion near the ground state of both axial motional modes, whilst maintaining coherence of the qubit. This is an essential ingredient in trapped-ion quantum computers. The isotope shifts are  sufficient to suppress decoherence and phase shifts of the memory qubit due to the cooling light which illuminates both ions. We measure the qubit coherence during 10 cycles of sideband cooling, finding a coherence loss of 3.3\% per cooling cycle. The natural limit of the method is $O(10^{-4})$ infidelity per cooling cycle.
\end{abstract}

\pacs{03.67.-a, 37.10.Ty, 37.10.Rs}


\maketitle

Trapped ions have been shown to have much promise for quantum
information processing (QIP). 
Multi-qubit quantum logic gates \cite{05Leibfried, 05Haffner,
03Leibfried, 03SchmidtKaler, 05Haljan, 06Home, 08:Benhelm}, high-fidelity operations \cite{08:Benhelm,08:Myerson},
teleportation and elementary algorithms
\cite{04Barrett, 04Riebe, 05Brickman, 05Chiaverini}
have been demonstrated. Scaling up from
these small scale demonstrations to algorithms involving large
numbers of gates, measurements and individual manipulations of a
large number of ions is a major challenge \cite{07Steane}. One
approach to this problem is to move the  ions themselves around a
large array of traps. This involves shuttling, separation, and
recombination of ion strings, which 
introduces heating \cite{02Rowe}. In
addition, ambient heating of ions has been widely observed in ion
traps, caused by fluctuations in the electric potential at the ion
\cite{00Turchette, 06DesLauriers,07Lucas}. All logic gate schemes
demonstrated thus far require the ions to be well within the
Lamb-Dicke regime for high-fidelity operation \cite{00Sorensen1,
03Leibfried}. Thus for a quantum processor involving ions in trap
arrays, the ability to cool ions near the ground state of motion
while preserving the logical information stored in them is
essential.

One approach to this problem is to cool sympathetically the qubit
ion, making use of its Coulomb interaction with another
``coolant'' ion stored in the same trap \cite{00Kielpinski,
01:Morigi}. Owing to the Coulomb interaction, the normal modes of
motion of the ions are shared, therefore by addressing laser cooling
only to the ``coolant'' ion we also cool the logical qubit ion. In
order that the light used for cooling does not decohere the qubit(s)
stored in the logic ion(s), it is necessary that it couples only
weakly to the internal state of the logic ion.

Sympathetic cooling of trapped ions near the motional ground
state has been reported in crystals of \caf--\caf \cite{01:Rohde},
$^{24}$Mg$^+$--$^{9}$Be$^+$ \cite{03:Barrett}
and $^{27}$Al$^+$--$^{9}$Be$^+$ \cite{07Rosenband}. 
In this paper we report near-ground state
sympathetic cooling using a pair of isotopes, \caf--\cafthree, in which we directly
measure the coherence of the qubit ion (by Ramsey interference)
while the cooling proceeds \footnote{In ref. \cite{07Rosenband}
there is evidence of coherence preservation during sympathetic Doppler
cooling near $\nbar \simeq 3$.}.

Consider a crystal of two ions: one coolant and one logic ion, the latter
storing a qubit in its internal state.
The goal is to cool one or more normal modes of motion from some initial
state $\rho_i$ to a cold final state $\rho_f$, without decohering the qubit.
Generally speaking, some form of resolved sideband cooling is needed, and
pulsed cooling is usually preferable because it is more readily optimized.
However, every process has some decohering effect.
We quantify this by a parameter $\epsilon$ defined as the drop in
interference fringe contrast observed in a Ramsey-type interference
experiment on the qubit, per cycle of pulsed sideband cooling of the ion pair.

In quantum computing one aims to suppress the infidelity per quantum
logic gate below some required level $\gamma$. Values in the range
$10^{-5} < \gamma < 10^{-3}$ are typically discussed \cite{07Steane}. Assuming the
logic gate is insensitive to the prepared motional state to lowest
but not higher orders, then the contribution of thermal motion to
the gate infidelity is of order $\gamma_T \simeq 0.3 \pi^2 \eta^4
\bar{n}(\bar{n}+1)$ where $\eta$ is the Lamb-Dicke parameter
and $n$ the motional quantum number \cite{00Sorensen1}. Since $\eta$ is typically of order $0.1$ this suggests
that $\bar{n}$ must be comfortably below 1, but it does not need to
be extremely small. Let $N$ be the number of cycles of sideband
cooling required to achieve this, then
$\gamma \ge \gamma_T + N \epsilon.$
$N$ is set by the gap between the desired motional state $\rho_f$ and
the state before cooling $\rho_i$, and by the ion masses. In a
computer, $\rho_i$ would be determined by electric field noise, and
by the precision of the ion transport and splitting/combining
operations in a trap array. We assume it would be roughly thermal,
with a mean excitation of order 1. In our experiments we study the
process in the relevant parameter regime by producing such
conditions in a single trap. Values in the region $1
\le N < 10$ are needed, and therefore the goal is $\epsilon \lesssim
10^{-4}$. This requires selective addressing, i.e.\ driving the
coolant ion through the cooling cycle, without exciting the internal
degrees of freedom of the logic ion.  $\epsilon$ is a function
primarily of the ratio $R$ of excitation rates of the two ions by
the applied laser beam(s). In pulsed sideband cooling, $R$ is the
number of photons scattered by the logic ion per cooling cycle.

Two approaches to attaining small $R$ have been discussed. The first
makes use of a tightly focussed laser beam to address one
ion only \cite{01:Rohde}. Assuming Gaussian optics and the same 
atomic coefficients for neighbouring ions,
one has $R \ge \exp(-2 s^2/w^2)$ where $s$ is the ion separation and $w$ the beam
waist. This approach involves very demanding technical constraints
which become greater at higher trap frequencies, since the ions are
closer together. It does however allow the logic and coolant ions to
be the same species, which has the advantage that the same
laser sources can be used for both the cooling and the control
of the logic ion.

The second approach is to use two different ion species.  Here
the cooling light couples weakly to the logic ion because
it is far (many nm) from resonance with transitions in this
ion. This approach has been used with a $^{24}$Mg$^+$--$^9$Be$^+$
crystal to achieve ground state cooling \cite{03:Barrett}.

We adopt a variant of the second approach, using two different isotopes of
the same element. This has previously
been used to Doppler cool ions \cite{02:Blinov}. There are some
advantages to having the two ions of similar mass \cite{00Kielpinski}, and
through the use of electro-optic modulators (EOMs) the same laser
sources can be used for control of both isotopes. However, off-resonant
coupling of cooling light to the logic ion is stronger than for the case of 
two different elements.

\begin{figure}
\centering
\includegraphics[width = 80mm]{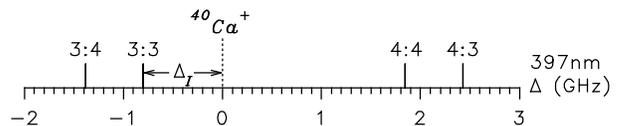}
\caption{Frequencies at zero magnetic field of the four
hyperfine components of the
S$_{1/2}\rightarrow$ P$_{1/2}$ transition in \cafthree,
relative to the same transition in \caf. The labels give
$F_{\rm S}$:$F_{\rm P}$, the total angular momentum quantum number
for the S (P) level. $\Delta_I$ in eq.~\protect\ref{R} is taken as
the smallest isotope shift in the set; the influence of the
other components is included via the factor $h$.}
\label{fig0}
\end{figure}

For the laser detunings used in our pulsed sideband cooling experiments,
$R$ is given approximately by
\be R = R\sub{rsb} + R_{\sigma} 
        \simeq g \frac{\pi \Gamma}{\eta \Delta}
         + h \left(\frac{\Gamma}{2 \Delta_I}\right)^2
\label{R} \ee 
where $\Gamma$ is the decay rate of the P$_{1/2}$
level, $\Delta$ is the detuning of the Raman sideband cooling laser
from the \caf S$_{1/2}\rightarrow$P$_{1/2}$ transition, $\Delta_I$
is the isotope shift (see Fig.~\ref{fig0}) and $g, h$ are
factors of order unity which take account of the angular momentum
coefficients and the sum over all the hyperfine components.

The two terms represent scattering during the red-sideband transition
and during the repump pulse respectively. Not all
photon scattering leads to decoherence of the qubit
\cite{05Ozeri, 06Ozeri}. However, for the parameter values
of our experiments the difference between the total scattering
rate and the inelastic scattering rate is small (we return to this
point below). The resulting decoherence from both sources is $\simeq R$.

In addition to photon scattering, which irreversibly decoheres the
qubit, the cooling beams also differentially light-shift the qubit
levels, resulting in a $\hat{\sigma}_z$ rotation, and they drive
various off-resonant Raman processes in the logic ion. The light-shift is
well-defined and can be taken into account or compensated in a
practical computer. The Raman processes
are hard to keep track of and therefore contribute to $\epsilon$; 
they can be kept small by a judicious choice
of magnetic field and laser polarization,
and by reducing the laser intensities.

In our experiments the logical qubit is stored in the \cafthree\ ``clock'' states
$\ket{\downarrow} \equiv \ket{S_{1/2},F=4,M_F=0}$ and
$\ket{\uparrow} \equiv \ket{S_{1/2},F=3,M_F=0}$. We
have observed long coherence times for these states \cite{07Lucas},
making this a good candidate for QIP. The \caf~ion is the coolant,
cooled by Doppler and Raman cooling via the 397~nm
4S$_{1/2}\rightarrow$ 4P$_{1/2}$ transition, hereafter designated
SP, with repump light on 3D$_{3/2} \rightarrow$ 4P$_{1/2}$. The SP
frequencies of \cafthree relative to \caf are given in Fig.~\ref{fig0}.

The ions are stored in a linear Paul trap largely as described in
\cite{00Barton}. The secular frequencies for a single trapped \caf\
ion are $\{\omega_r, \omega_z\} \simeq  2 \pi \times \{700,500\}$
kHz for radial and axial motion respectively. A two ion crystal of
\caf--\cafthree exhibits radial oscillation near $\omega_r$ and two
modes of axial motion. In the lower (higher) frequency axial mode, the
oscillation of the two ions is nearly in-phase (out-of-phase).
The axial mode frequencies are $\omega_{\rm in} =
0.98\ \omega_z$ and $\omega_{\rm out} = 1.70\ \omega_z$.

A $0.17\,$mT magnetic field splits the two ground
states of \caf by $\omega_0 \simeq 2\pi \times 4.8$~MHz, and removes
the Zeeman degeneracy of \cafthree. Fluorescence
from each ion is observed by turning on its respective 397~nm
Doppler cooling beam. The \cafthree\ 397~nm beam
is tuned to the $F=4 \rightarrow 4$ hyperfine component of SP
and an EOM at 3.220 GHz provides a sideband to repump on 
$F=3 \rightarrow 4$.
Population in the D levels is repumped for both ions
by lasers at 866~nm, 850~nm and 854~nm.

Three stages of cooling (of \caf) are required to reach the ground
state of motion. The first is Doppler cooling using the 397~nm beam,
detuned half a linewidth below SP. The second stage is continuous
Raman sideband cooling using two beams derived from the same laser,
detuned 130~MHz below SP. The final stage is pulsed Raman sideband
cooling. In each cooling cycle, first a Raman red sideband
$\pi$-pulse transfers population from S$_{1/2}$ $\ket{M_J = -1/2, n}
\rightarrow \ket{M_J = +1/2, n - 1}$ (mean duration $15\,\mu$s),
then a $\tau_{\sigma}=10\,\mu$s repump pulse, resonant with SP and polarized
$\sigma^-$, is used to optically pump from $M_J = +1/2$ to $M_J =
-1/2$ via P$_{1/2}$.

The Raman sideband pulses are implemented using two beams derived
from the same laser, detuned by $\Delta = 2 \pi \times$30~GHz above
SP and with a frequency difference $\delta$ introduced by AOMs. One
beam is directed along the magnetic field and is vertically
polarized, the other is at 60$^\circ$ to the magnetic field and is
horizontally polarized. The difference wavevector is along the axis
of the trap. The repump beam is directed along the magnetic field.

\begin{figure}
\centering
\includegraphics[width = 0.37\textwidth]{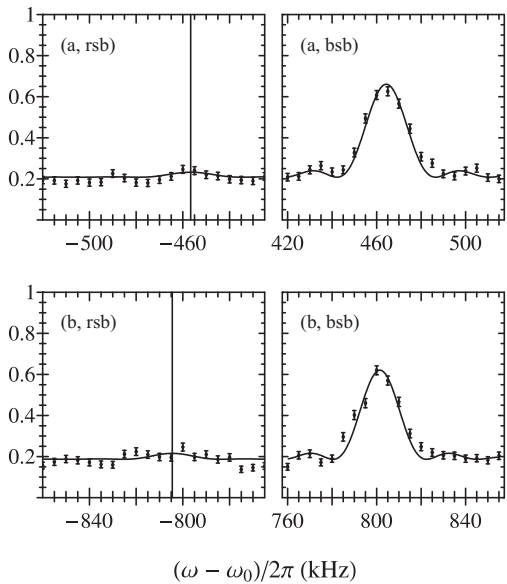}
\caption{Results of scanning a 24~$\mu$s Raman
probe pulse over the red and blue sidebands of a) the in-phase mode
and b) the out-of-phase mode, after cooling. The data shows the fraction of
times in 500 repetitions that no $^{40}$Ca$^+$ fluorescence was observed
after a detection pulse. It is
fitted with sinc functions, and from the ratio of the fitted
amplitudes we obtain $\bar{n}_{\rm in} = 0.06(3)$, $\bar{n}_{\rm
out}  = 0.07(5)$. Vertical lines indicate the (known)
frequencies of the red sidebands.}
\label{fig1}
\end{figure}

After cooling both axial modes, motional temperatures were inferred
from sideband observations \cite{89Diedrich}. 
With a fixed-length probe pulse, $\delta$ is
scanned and the excitation probability detected, see 
Fig.~\ref{fig1}. For a thermal state in the Lamb-Dicke regime, the
ratio $r$ of amplitudes of the red and blue sidebands gives the mean
vibrational quantum number for the mode: $\bar{n} = r/(1 - r)$. For
the data shown, we find $\bar{n}_{\rm in} = 0.06(3)$ and
$\bar{n}_{\rm out} = 0.07(5)$ for the in and out-of-phase modes
respectively.

Next we describe the experimental observation of qubit coherence
during sympathetic sideband cooling of the out-of-phase mode.
The results extend readily to cooling multiple modes. We measure the
mode temperature as above, and
we measure the qubit coherence by Ramsey interferometry.
A given experimental sequence consists of pre-cooling to $\bar{n} \sim 0.6$
(of the order to be expected after controlled transport in an ion trap computer)
then a standard Ramsey sequence, with
one or more sideband cooling cycles {\em inserted in the gap} between
the Ramsey $\pi/2$ pulses, followed by detection of either the qubit state
or the motional state. One complete
measurement of the pair $(\bar{n}_{\rm out},\,\epsilon)$ involves
40 data points (20 each for scans of final sideband strength and
Ramsey frequency), where each data point is obtained from
500 repetitions of a given experimental sequence.
The Ramsey fringes are fitted with a sine function and compared
to a control experiment with no sideband cooling.

Qubit state preparation is by turning off the EOM in the 397~nm beam.
This results in
optical pumping to S$_{1/2}$ $F = 3$, with near uniform filling of
the Zeeman sublevels, and therefore 15\% preparation of
$\ket{\uparrow}$. Resonant Rabi flopping of the qubit is driven by
3.226~GHz microwaves applied to one of the trap electrodes. Readout
is accomplished using a short pulse of circularly
polarized 393~nm light to selectively transfer population from $F =
4$ into the D$_{5/2}$ ``shelf''. The shelving probability is
$0.90$, $0.002$ for $\ket{\downarrow}, $\ket{\uparrow} respectively.
The Doppler cooling lasers are then turned on, and the presence
of fluorescence indicates whether the ion was shelved.

\begin{figure}
\includegraphics[width = 0.45\textwidth]{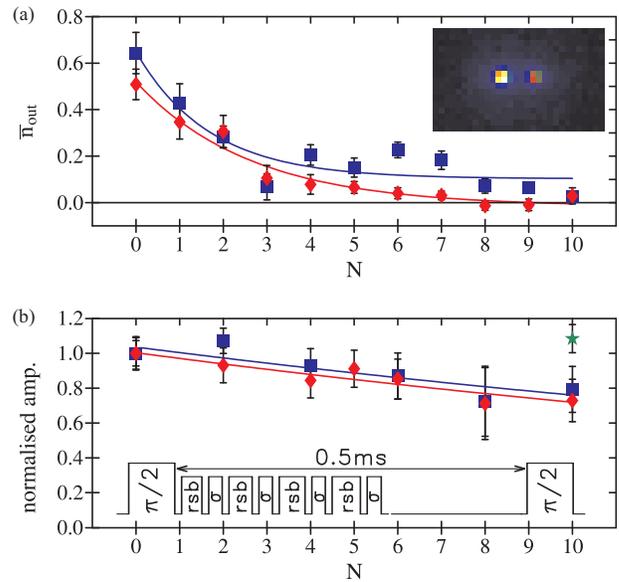}
\centering 
\caption{(a) $\bar{n}_{\rm out}$ as a function of
the number of cycles $N$ of pulsed Raman sideband cooling. Each point
is obtained from data similar to that shown in Fig.~\protect\ref{fig1}. (b)
Ramsey fringe amplitude, normalised to a control experiment
with no sideband cooling cycles.
Two data sets are shown ($\blacksquare$, $\blacklozenge$); 
the results are fitted with exponential decay curves. Insets: two-ion fluorescence
image (a) and pulse sequence (b).
Also shown is a single point ($\star$) from an experiment in which only 
the 10 repump $\sigma$ pulses were applied. The amplitude of this fringe pattern
relative to the control experiment is $1.08(9)$.} 
\label{fig2}
\end{figure}

Our results are shown in Fig.~\ref{fig2}. 
Mean vibrational quantum number and Ramsey fringe amplitude
are shown as a function of $N$
for two sets of data. The fringe amplitude was fitted with an exponential function
with floated decay constant, giving a loss of coherence per cooling cycle
$\epsilon = 0.033(4)$. 
Half of this error is due to scattering (eq. \ref{R}); the other half agrees
with the estimated contribution of off-resonant Raman processes driven by the 
cooling beams. 

Much lower values of $\epsilon$ can be reached.
Our experimental parameter values were such that
$R$ is dominated by photon scattering during the sideband pulse $R\sub{rsb}$. We performed two
further experimental tests to establish this and to explore what the natural
limits of the method are.

\begin{figure}
\includegraphics[width = 86mm]{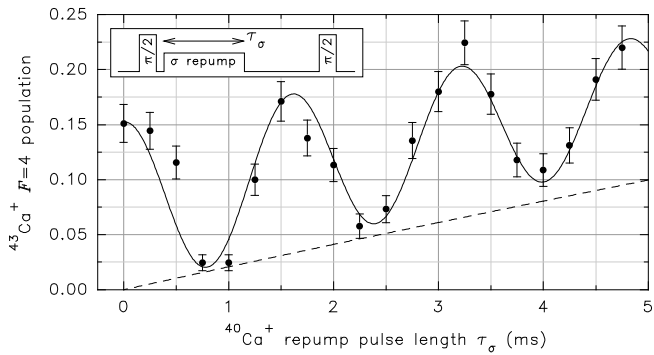}
\centering 
\caption{The population of the \cafthree\ S$_{1/2}^{F=4}$ level as a function
of the duration of a \caf\ repump pulse inserted between two $\pi/2$
pulses on the ``clock'' qubit transition. The initial population of
S$_{1/2}^{F=4,M_F=0}$ is $\simeq 0.15$.} 
\label{fig3}
\end{figure}

Fig.~\ref{fig2}(b) shows a further data point from an
experiment performed with only the 10 repump pulses in the gap
between the $\pi/2$ pulses. From this we calculate an upper bound
$\epsilon < 7 \times 10^{-3}$ (with no cooling)
at 95\% confidence level.
In the second test, a single pulse from the repump laser was turned on in the 5 ms gap
of a Ramsey experiment. The observed population of the $F = 4$
hyperfine level as a function of the repump pulse duration is shown in Fig.~\ref{fig3}. 
The scan shows three major effects, for which values are
obtained by fitting
\[
P(F = 4) = \left(1 - e^{- \alpha t}\right) + A \left[ 1+e^{-\beta t}\cos(\Delta_q t) \right]. 
\] 
These are: 
a rising baseline at rate $\alpha = 21(3)$~s$^{-1}$ due to optical pumping
from the `spectator' $F = 3, M_F \neq 0$ states to $F = 4$,
oscillations of amplitude $A=0.076(8)$ and frequency $\Delta_q = 2 \pi \times 623(7)$ Hz
due to the differential light shift of the qubit levels, and decaying amplitude
of these oscillations at $\beta = 60(50)$~s$^{-1}$. 
$A$ indicates that 15\% of the population was prepared in the clock state.
$\Delta_q$ does not decohere the qubit, but it implies a phase shift
of 39~mrad per 10~$\mu$s repump pulse which must be taken into account.
From $\Delta_q$ we infer the laser intensity was $10.6$ W/m$^2$.
Putting this into a rate equation model of the optical pumping gave an expected
transfer to $F=4$ in agreement with the observed baseline slope $\alpha$.
The decaying amplitude of the oscillations represents decoherence but it
is only just observable; the data suggest an upper bound $\beta < 150$~s$^{-1}$
and hence $\epsilon = \beta \tau_{\sigma} < 1.5 \times 10^{-3}$. The baseline
slope $\alpha$ directly measures the $F=3 \rightarrow 4$ pumping rate for approximately
uniform filling of the Zeeman sublevels of S$_{1/2}$, $F=3$. The scattering rate $R_{\sigma}$
for an ion 100\% in the clock manifold ($\ket{\uparrow},\, \ket{\downarrow}$), the second term in 
eq.~(\ref{R}), can be inferred from this using the known relative component
strengths. We obtain $R_{\sigma} \simeq 2 \alpha \tau_{\sigma} = 4(1)\times 10^{-4}$~s$^{-1}$. 
This is our best estimate of the contribution to $\epsilon$ from the repump process
in our cooling experiments.


To conclude, we have observed sympathetic sideband cooling to the ground state 
of motion ($>90$\% occupancy for two modes),
and directly demonstrated that laser-cooling of a trapped ion 
quantum register can be implemented 
in between coherent qubit rotations with small (3\%) cost in fidelity. 
Smaller error rates are in principle available by increasing the power and detuning of the Raman sideband cooling laser:
this reduces the first term in eq.~(\ref{R}). Eventually one reaches the regime where Rayleigh
elastic scattering dominates, and for our chosen transition in Ca$^+$ the contribution to $\epsilon$ is
$O(10^{-4})$ \cite{06Ozeri}. The second term in (\ref{R}) is
calculated to be $1 \times 10^{-4}$ when 3 photons are scattered from the \caf~ion.
Extraneous Raman processes in the logic ion can also be reduced to this
level by using pure $\sigma$- and $\pi$-polarized cooling beams and a sideband Rabi
frequency smaller by a factor 2.
Supposing one coolant ion is used to cool two logic ions, one finds the error from Raman
scattering associated with one cooling pulse is similar to that
associated with one 2-qubit logic gate pulse.
To achieve even lower error rates, one could use a transition with smaller width
and larger isotope shifts, such as to a metastable D state in Ca$^+$, 
either for cooling or for logic gates or both \cite{01:Rohde,08:Benhelm}. 


This work was supported by IARPA (contract W911NF-05-1-0297), the EPSRC (QIP
IRC), the European Commission (FET networks CONQUEST, SCALA)
and the Royal Society.




\end{document}